\documentclass[prl,twocolumn,showpacs,floatfix]{revtex4}
\usepackage{epsfig}
\usepackage{color}

\begin{document}

\title{The Complete Jamming Landscape of Confined Hard Discs}

\author{S. S. Ashwin}\thanks{Both authors contributed equally to this work}
\affiliation{Department of Chemistry, University of Saskatchewan,
Saskatoon, Saskatchewan, S7N 5C9}

\author{Richard K. Bowles}\thanks{Corresponding Author: richard.bowles@usask.ca}
\affiliation{Department of Chemistry, University of Saskatchewan,
Saskatoon, Saskatchewan, S7N 5C9}
\email{richard.bowles@usask.ca}

\begin{abstract}
An exact description of the complete jamming landscape is developed for a system of hard discs of diameter $\sigma$, confined between two lines separated by a distance $1+\sqrt{3/4} < H/\sigma < 2$. By considering all possible local packing arrangements, the generalized ensemble partition function of jammed states is obtained using the transfer matrix method, which allows us to calculate the configurational entropy and the equation of state for the packings. Exploring the relationship between structural order and packing density, we find that the geometric frustration between local packing environments plays an important role in determining the density distribution of jammed states and that structural ``randomness" is a non-monotonic function of packing density. Molecular dynamics simulations show that the properties of the equilibrium liquid are closely related to those of the landscape.
\end{abstract}


\maketitle

The properties of a wide variety of materials, including liquids, glasses, crystals and granular materials, depend on the way the particles pack.  Bernal~\cite{ber60} originally used random packings of ball bearings to study the structure of liquids and coined the term random close packing (RCP) to describe the most dense random arrangement of spheres. Subsequently, the potential energy landscape~\cite{gold69} (PEL) has become an important paradigm used to describe the role of particle packing in both the thermodynamics and dynamics of many of these systems~\cite{still95,deb01}. In this approach, each configuration of the liquid is represented as a point in the high-dimensional, $N$-body, potential energy function of the system that can be uniquely mapped to the closest mechanically stable packing or inherent structure~\cite{stil64}. For a system with a soft potential, an inherent structure represents a local potential energy minimum. For a hard particle system, an inherent structure is a collectively jammed packing~\cite{don05,don05b} and a local density maximum, which gives rise to the corresponding jamming landscape (JL). The configurations that map to the same inherent structure can be grouped together into a local basin of attraction and the properties of the liquid can be described in terms of the number of inherent structures and the motion of the system through the resulting landscape of basins and saddle points. 


It has been suggested that RCP might be related to an ideal glass state. In the context of the JL, the configurational entropy of a hard sphere system is defined as $S_c=\ln \Omega(\phi_j)$, where $\Omega(\phi_j)$ is the number of collectively jammed states with an occupied volume fraction $\phi_j$, so the ideal glass transition corresponds to a density where the metastable liquid would become trapped in a single unique basin and $S_c\rightarrow 0$. However, obtaining a detailed description of the JL remains a considerable challenge. Computer simulation has been used extensively to study packing, but different protocols often lead to different conclusions regarding the density distribution of inherent structures for both hard disc mixtures ~\cite{spe01,oh5,tor06} and hard spheres~\cite{oh2,ani08}. A recent study~\cite{tor00} of jammed packings also raised questions concerning the relationship between the structure of a packing and its density. In particular, it has been suggested that the  RCP should be replaced by a maximally random jammed (MRJ) state that is more rigorously defined with respect to a set of order parameters. The only exact analytical results available for the entire JL are for one-dimensional~\cite{rkb00} or quasi-one-dimensional~\cite{rkb06} hard particle systems where particles can only interact with a single neighbor on each side. A recent mean field theory~\cite{mak08}, which finds hard sphere packings ranging from $\phi_j=0.536-0.635$, represents one of the few analytical results describing the JL in higher dimensions.

The goal of the present work is to obtain an exact description of the JL for a system so that we can explore the relationship between packing structure and density. To this end, we study a system of two-dimensional (2d) hard discs of diameter $\sigma$, confined between two hard walls separated by a distance of $1+\sqrt{3/4} < H/\sigma < 2$. In 2d, a particle is locally jammed if it has at least three rigid contacts that are not all in the same semicircle. However, local jamming of all the particles is a necessary but not sufficient condition for collective jamming because the concerted motion of a number of particles can lead to a collapse of the structure~\cite{don05}. By confining the discs to a channel with $H/\sigma<2$, we prevent the particles from passing each other and eliminate the possibility of collective rearrangements. The number of structures in which all the particles are locally jammed is then equal to the number of collectively jammed inherent structures. For $H/\sigma<1+\sqrt{3/4}$, discs can only contact their nearest neighbors, which only allows two local particle arrangements that jam and gives rise to a binomial-type density distribution of inherent structures~\cite{rkb06}. 

In the range of channel diameters studied here, the discs can contact both their first and second nearest neighbors, leading to a significant increase in the number and structural diversity of possible locally jammed environments. The local packings are identified using a heuristic algorithm that generates new packings by making perturbative moves of individual particles, or groups of particles, on the known jammed structures for a system with $H/\sigma = 1+\surd{3}/4$. Our rationale for the algorithm is based on the observation that the particle arrangements for $1+\sqrt{3/4} < H/\sigma < 2$ must collapse continuously back to one of the jammed structures at $1+\surd{3}/4$ because there is no change in the number of possible contacts a disc can have over this range of channel diameters.

\begin{figure}[h]
\hbox to \hsize{\epsfxsize=1.0\hsize\hfil\epsfbox{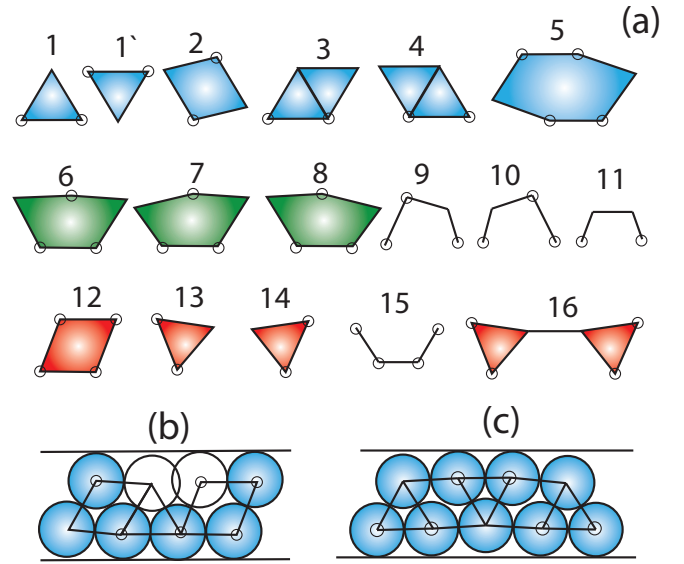}}
\caption{(a) The set of tiles that represent the local jam packed configurations. The vertices represent disc centres and the solid lines join the centres of two contacting discs. The small open circles identify those discs that contact the wall. An additional 16 tiles are generated by reflecting each tile in the plane of symmetry that runs through the axial center line of the channel, e.g. tile $1^{\prime}$ is generated from tile 1. (b) A configuration constructed from tiles $2^{\prime} - 1 - 12$. A tile is added to the sequence by sharing the first two discs with the previous tile, with the exception of tile 14, which only shares one disc, and tiles 6-8, which may either simply contact the previous tile, or share discs, depending on the nature of the neighbouring tile. In our example, tile 1 is compatible with $2^{\prime}$ ($C_{2^{\prime},1}=1$) but tile 12 is incompatible with 1 because it causes particle overlap between the unfilled discs  ($C_{1,12}=0$).  Other incompatibilities result in discs remaining unjammed. (c) Most dense packing.}
\label{fig1:tiles}
\end{figure}

All the local packing environments are then mapped onto a set of 32 tiles (Fig.~\ref{fig1:tiles}a) that can be combined in a sequence, from left to right, such that the local jamming conditions for each particle can be achieved by just considering the neigboring tiles. This allows us to use the transfer matrix method to construct the exact partition function for all the jammed states. Incompatibilities exist between some of the tiles in the sense that they cannot form a left-right pair that results in a valid, jammed packing of the particles (Fig.~\ref{fig1:tiles}b). This results in three groups of tiles; a set of high density tiles (tiles 1-5 in Fig.~\ref{fig1:tiles}a), so called because they appear in the high density inherent structures; a set of low density tiles (tiles 12-16) and a set of interface tiles (6-11). There is no direct compatibility between the low and high density tiles except that interface tiles are compatible with some members of both other sets, so it is possible to mix the groups within a single packing. The separation in the packing compatibility arises because the two important length scales in the system, $\sigma$ and $H$, are incongruent. Tiles $12 - 16$ all contain a disc-disc contact that spans the width of the channel at an angle that cannot jam the particles in the other set of tiles.

We begin to construct the partition function by defining the $32\times32$ transfer matrix, $M$,  with matrix elements
\begin{equation}
M_{ij}=C_{ij}K^{\beta\mu n_{ij}}L^{-\beta P_{LS}H l_{ij}}\mbox{,}\\
\label{eq:tm}
\end{equation}
where $C_{ij}=1$ if tiles $i$ and $j$ are compatible but is zero otherwise. $n_{ij}$ and $l_{ij}$ are the number of particles and length associated with the addition of tile $j$ to the right of tile $i$. $\mu$ is the chemical potential and $P_{LS}$ is the external longitudinal landscape pressure applied to the ends of the channel, rather than the pressure related to the internal vibration of the particles which must be zero for the jammed states. $\beta=1/kT$, where $T$ is the temperature and $k$ Boltzmann's constant. 
The definition of a temperature for a system of jammed states is of considerable general interest~\cite{temp}. However, here we note that the configurational integral of a hard particle system is independent of $T$ so we are free to simply treat temperature as a parameter conjugate to the entropy. Furthermore, an analysis of the transfer matrix shows that the number of states with a given $N$ and $V$ appear as coefficients of the polynomial matrix elements of $M^{N_T}$ that can be extracted by taking the appropriate derivative with respect to $T$ without necessarily providing a clear thermodynamic definition of temperature.

Taking $M^{N_T}$ gives all the possible jammed packing arrangements that can be formed with $N_T$ tiles. Both the number of particles in a packing, $N$, and its volume, $V$, fluctuate between different packings, so we use the generalized ensemble partition function~\cite{hill}, which can be expressed
\begin{equation}
\Gamma(P_{LS},\mu,T)=\sum_{N_T=0}^{\infty}\sum_{i=1}^{32}\lambda_i^{N_T}=\sum_{i=1}^{32}\frac{1}{1-\lambda_i}\mbox{ ,}
\label{eq:pf}
\end{equation}
where $\lambda_i$ is the $i^{th}$ eigenvalue of $M$. {
The equilibrium condition is obtained by finding $P^*_{LS}$ and $\mu^*$ such that $\Gamma(P^*_{LS},\mu^*,T)=1$. In the thermodynamic limit, the equilibrium properties of the ensemble averages are overwhelmingly determined by packings of a given $N$ and $V$, which are obtained from,
\begin{eqnarray}
\bar{N} &=&\partial_{\mu} \ln\Gamma(P_{LS},\mu,T) \mid_{(P_{LS},\mu)\rightarrow(P^*_{LS},\mu^*)}\\
\bar{V} &=&\partial_{P_{LS}} \ln\Gamma(P_{LS},\mu,T) \mid_{(P_{LS},\mu)\rightarrow(P^*_{LS},\mu^*)}\mbox{ ,}
\label{eq:nv}
\end{eqnarray}
and the configurational entropy,
\begin{equation}
S_c =kT\partial_{T} \ln\Gamma(P_{LS},\mu,T)\mid_{(P_{LS},\mu)\rightarrow(P^*_{LS},\mu^*)}\mbox{ ,}\\
\label{eq:sc}
\end{equation}
is the logarithm of the number of such packings. We also calculate the fraction of particles, $f_{ij}$, belonging to interfaces between tiles $i$ and $j$, as follows: if an additional factor $\gamma$ is associated with matrix element $M_{ij}=\exp(-\beta P_{LS}H L_{ij}+\mu\beta\gamma N_{ij})$ then,
\begin{equation}
f_{ij}=(1/\mu \bar{N}) \partial_{\gamma} \ln\Gamma(P_{LS},\mu,T,\gamma) |_{(P_{LS},\mu)\rightarrow(P^*_{LS},\mu^*)}\mbox{.}
\label{eq:fij}
\end{equation}

All results reported here are given for a channel width of $H=1.95\sigma$. Fig.~\ref{fig2:sc} shows $S_c/Nk$, which is independent of $T$, as a function of the $\phi_j$ over the entire range of packing densities, except at the very extremes where it was not numerically possible to find solutions where $P_{LS}$ and $\mu$ becoming extremely large. Both the most dense and least dense states are non-degenerate, except with respect to identical structures obtained by a reflection along the axis of the channel. The most dense state, with $\phi_j\approx 0.80743$, consists of a repeated tile sequence of $-1-2-1^{\prime}-2^{\prime}-$ (Fig.~\ref{fig1:tiles}c), while the least dense packing is made from alternating $-15-15^{\prime}-$ tiles, giving $\phi_j\approx 0.613$. The maximum in $S_c$ occurs at $\phi_j=0.712$, which also corresponds to the density sampled by the system at $P_{LS}=0$ (see equation of state (EOS) in Fig.~\ref{fig3}a). This is consistent with earlier studies that found configurations of the ideal gas were related to the maximum in the configurational entropy. The lower density packings are sampled at negative pressures, which are non physical in an equilibrium hard particle fluid, while the system samples deeper basins with increasing $P_{LS}$.  Above $\phi_j=0.806$, we see a smooth, but rapid change in the EOS as $S_c$ dramatically decreases.

\begin{figure}[t]
\hbox to \hsize{\epsfxsize=1.0\hsize\hfil\epsfbox{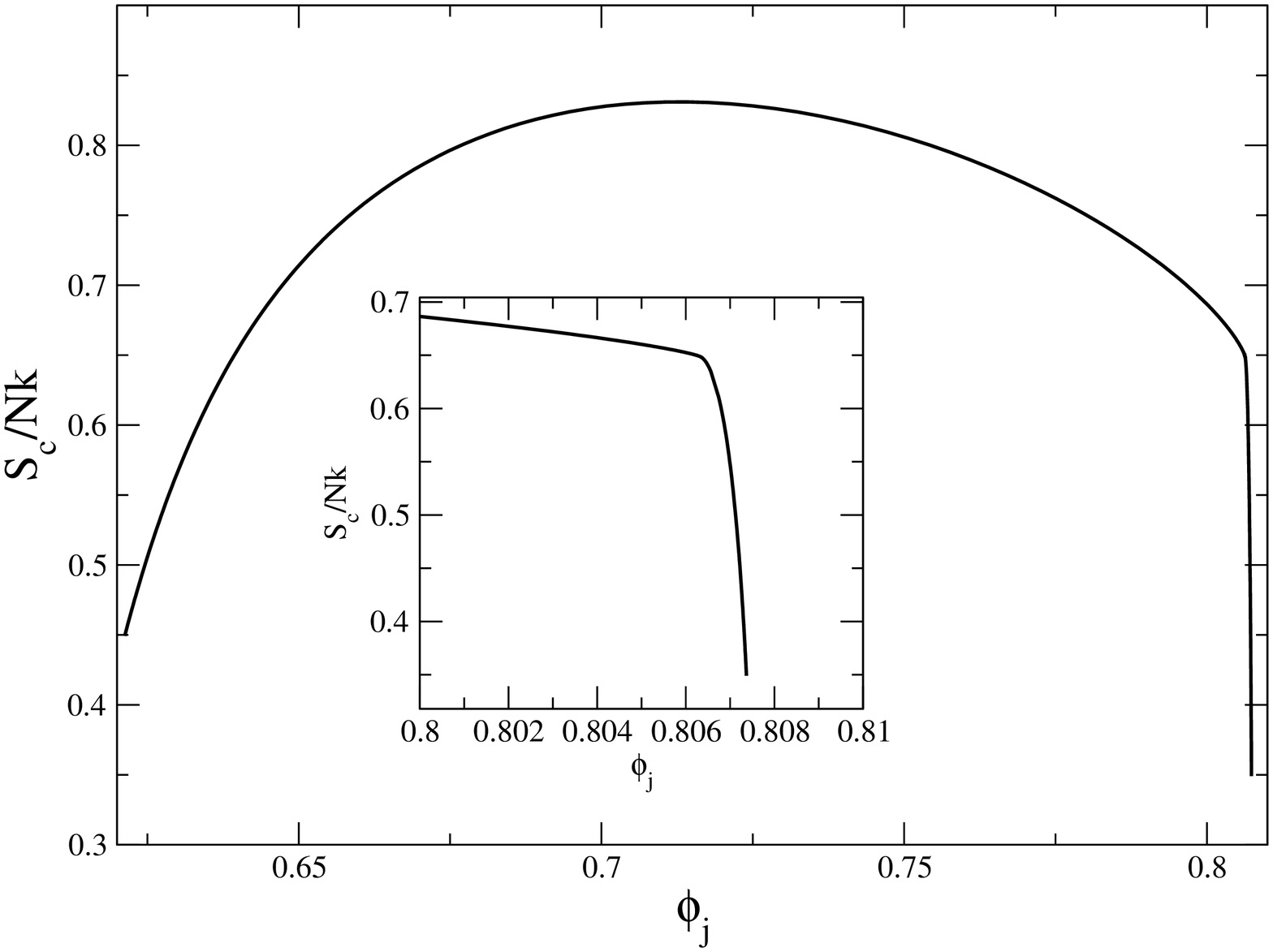}}
\caption{Configurational entropy, $S_c$ vs $\phi_j$. Insert: Enlarged high density region.}
\label{fig2:sc}
\end{figure}

\begin{figure}[t]
\hbox to \hsize{\epsfxsize=1.0\hsize\hfil\epsfbox{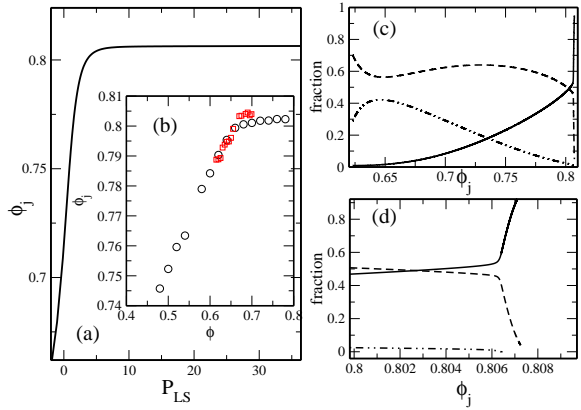}}
\caption{(a) $\phi_j$ as a function of $P_{LS}$. (b) $\phi_j$ as a function of the equilibrium fluid $\phi$ obtained from simulation studies with $N=996$ (circles) and $N=7998$ (squares) particles. (c) The fractions of particles in tile interfaces between high density tiles, $f_{hd}=\sum_{i,j=1}^5 f_{ij}$ (solid line), low density tiles, $f_{ld}=\sum_{i,j=12}^{16} f_{ij}$ (dashed line) and interface tiles, $f_{int}=\sum_{i=6,j=1}^{11,16} f_{ij}$ (dot-dashed line). (d) high density enlargement of (c).}
\label{fig3}
\end{figure}

The structure of the packings is examined by measuring the fraction of particles associated with the high density tiles, $f_{hd}$, low density tiles, $f_{ld}$, and interface tiles, $f_{int}$, as a function of $\phi_j$ (Figs.~\ref{fig3}c and \ref{fig3}d). The system can increase $\phi_j$ in two ways: by replacing the more open tiles of a set with more dense tiles from the same set, i.e. replacing tile 15 with tile 9; or by replacing low density tiles with high density tiles. Fig.~\ref{fig3}c shows that both these processes are occurring over the entire density range, but we also notice that $f_{int}$ is always decreasing above $\phi_j=0.64$. This implies that the packings are becoming more structurally heterogeneous and contain clusters of high and low density tiles as their ability to mix is reduced. By $\phi_j=0.806$, most of the low density tiles remaining are of type 9, so that the only way to increase the density further is to eliminate the low density tiles altogether, which results in a rapid structural crossover and the decrease in $S_c$.

Structural randomness would imply that it is equally probable to find any given packing arrangement, on all possible length scales. Consequently, the function, $\Psi=(1/d_{max})\sum_{d,i,j} p_{ij}(d)\ln p_{ij}(d)$,
where $p_{ij}(d)$ is the probability of finding tile $j$ a distance $d$ from tile $i$ and the sums are over all possible tile pairs and separations, should be a maximum for the most random states. $d_{max}$ is the maximum separation studied. $\Psi$ is a measure of the average randomness of the packings at a given $\phi_j$ and is likely to provide a good description of the liquid structure as a function of density. In contrast, the MRJ state focuses on the randomness of a particular structure. Fig.~\ref{fig4:rand} shows the unexpected result that randomness is not a monotonic function of $\phi_j$. The global maximum in $\Psi$ occurs at $\phi_j\approx0.66$ and it initially decreases with increasing $\phi_j$ as a result of the clustering of low and high density tiles described above. The elimination of the low density tiles then allows greater mixing between the high density tiles, causing $\Psi$ to increase again to its second maximum at $\phi_j\approx 0.8071$.

We use event driven, molecular dynamics (MD) simulations~\cite{ald60} to understand how the thermodynamics of the equilibrium fluid are related to the landscape. The system is decompressed in small increments from $\phi=0.8072$, using the most dense packing as the starting configuration, to $\phi=0.1$. At each $\phi$ studied, $100000N$ collisions are used to reach equilibrium and the data are collected over the next $25000N$ collisions, where $N$ is the number of discs in the simulation. 200 independent configurations from each $\phi$ are then compressed until they are jammed using a modified version of the Lubachevsky and Stillinger~\cite{lscomp} (LS) algorithm that ensures $H/\sigma$ remains constant as the diameter of the discs is changed ($L$ fixed).  A compression rate $\partial \sigma/\partial t=0.04$ is used. This process is not an ``infinitely" fast quench and the system is able to relax via MD while it is compressed, but faster quench rates result in poorly jammed states with loose discs. Fig.~\ref{fig3}b shows that the average $\phi_j$, obtained by quenching the equilibrium liquid from $\phi$ for systems with $N=996$ and $N=7998$ discs, displays a dramatic change in its $\phi$ dependence above $\phi\approx 0.66$. This change is also apparent in the EOS (Fig.~\ref{fig3}a) for the jammed system obtained directly from the partition function. While the EOS always varies smoothly, signifying that there is just one thermodynamic fluid phase at all densities, the rapid change in slope of $\phi_j$ coincides with the structural crossover observed in Fig.~\ref{fig3}d.


\begin{figure}[t]
\hbox to \hsize{\epsfxsize=1.0\hsize\hfil\epsfbox{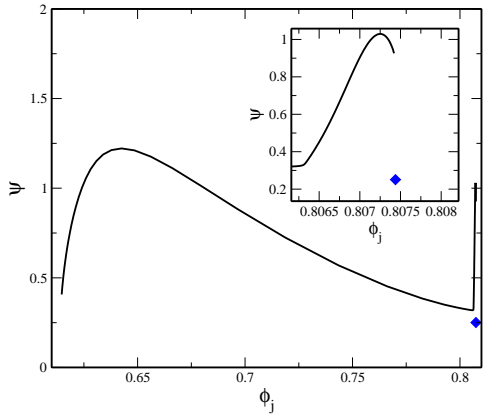}}
\caption{$\Psi$ as a function of $\phi_j$ obtained at $T=0.15$. The diamond represent $\Psi$ for the most dense state, obtained directly from its structure. Insert: An enlargement of the high density region.} 
\label{fig4:rand}
\end{figure}

More generally, our work highlights the importance of understanding the role of local packing environments and geometric frustration in determining the properties of hard particle packings, and its relationship to packing randomness. Anikeenko {\it et al.}~\cite{ani08,ani07} also observe competition between the ``quasiperfect'' tetrahedra of the disordered packings and the crystalline arrangements of hard spheres. This suggests frustration may be a generic feature of packings and that a tiling approach may be useful in enumerating inherent structures~\cite{ash}. We also find that $S_c$ is exponential in the number of particles over the entire density range of $\phi_j$, which is consistent with Donev {\it et al.}~\cite{tor06}, who used computer simulation show that this is the case for binary mixtures of hard discs in the bulk. Finally, the complete knowledge of the JL for a system, along with an understanding of local packing, will allow us to directly test the relationships between the landscape, kinetically facilitated dynamics~\cite{gar09}, geometric frustration~\cite{tar05}, and the glassy dynamics of liquids.

We would like to thank P. H. Poole, I. Saika-Voivod, C. E. Soteros and R. J. Spiteri for discussions. This work was supported by NSERC and CFI.


\end{document}